\documentclass[10pt]{article}
\usepackage{amsfonts}
\usepackage{graphicx,amssymb,mathrsfs,amsmath}
 \newtheorem{theorem}{Theorem}
 \newtheorem{lemma}{Lemma}
 \newtheorem{corollary}{Corollary}
 
 \newtheorem{definition}{Definition}
 \newtheorem{algorithm}{Algorithm}
 \newtheorem{remark}{Remark}
\begin{document}

\title{Finding Exact Minimal Polynomial  by Approximations
\thanks{The work is partially supported by China 973 Project NKBRPC-2004CB318003,
the Knowledge Innovation Program of the Chinese Academy of Sciences
KJCX2-YW-S02, and the National Natural Science Foundation of
China(Grant NO.10771205)}}
\author{
Xiaolin Qin \ Yong Feng \ Jingwei Chen  \ \ Jingzhong Zhang   \\
Laboratory for Automated Reasoning and Programming\\
Chengdu Institute of Computer Applications\\
Chinese Academy of Sciences\\
610041 Chengdu, P. R. China\\
Laboratory of Computer Reasoning and Trustworthy Computation\\
University of Electronic Science and Technology of China\\
610054 Chengdu, P. R. China \\
E-mail: qinxl811028@163.com, yongfeng@casit.ac.cn }
\date{}
\maketitle
\begin{abstract}
We present a new algorithm for reconstructing an exact algebraic
number from its approximate value using an improved parameterized
integer relation construction method. Our result is almost
consistent with the existence of error controlling on obtaining an
exact rational number from its approximation. The algorithm is
applicable for finding exact minimal polynomial by its approximate
root. This also enables us to provide an efficient method of
converting the rational approximation representation to the minimal
polynomial representation, and devise a simple algorithm to factor
multivariate polynomials with rational coefficients. \\
Compared with other methods, this method has the numerical
computation advantage of high efficiency. The experimental results
show that the method is more efficient than \emph{identify} in
\emph{Maple} 11 for obtaining an exact algebraic number from its
approximation. In this paper, we completely implement how to obtain
exact results by numerical approximate computations.

{\bf keywords: } Algebraic number, Numerical approximate
computation, Symbolic-numerical computation, Integer relation
algorithm, Minimal polynomial.
\end{abstract}
\section{Introduction}
\par Symbolic computations are principally exact and stable. However,
they have the disadvantage of intermediate expression swell.
Numerical approximate computations can solve large and complex
problems fast, whereas only give approximate results. The growing
demand for speed, accuracy and reliability in mathematical computing
has accelerated the process of blurring the distinction between two
areas of research that were previously quite separate. Therefore,
algorithms that combine ideas from symbolic and numeric computations
have been of increasing interest in recent two decades. Symbolic
computations is for sake of speed by intermediate use of
floating-point arithmetic. The work reported in
\cite{CG2001,CG2000,HS2000,SSM1991,CAW2002,FZQY2008} studies
recovery of approximate value from numerical intermediate results. A
somewhat related topic are algorithms that obtain the exact
factorization of an exact input polynomial by use of floating point
arithmetic in a practically efficient technique
\cite{CG2006,CFQZ2008}. In the meantime, symbolic methods are
applied in the field of numerical computations for ill-conditioned
problems \cite{S1985,C1966,B1971}. The main goal of hybrid
symbolic-numeric computation is to extend the domain of efficiently
solvable problems. However, there is a gap between approximate
computations and exact results\cite{YZH1992}.
\par We consider the following question: Suppose we are given an approximate
root of an unknown polynomial with integral coefficients and a bound
on the degree and size of the coefficients of the polynomial. Is it
possible to infer the polynomial and its exact root? The question
was raised by Manuel Blum in Theoretical Cryptography, and Jingzhong
Zhang in Automated Reasoning, respectively. Kannan \emph{et al}
answered the question in \cite{KLL1988}. However, their technique is
based on the Lenstra-Lenstra-Lovasz(LLL) lattice reduction
algorithm, which is quite unstable in numerical computations. The
function \emph{MinimalPolynomial} in \emph{maple}, which finds
minimal polynomial for an approximate root, was implemented using
the same technique. In this paper, we present a new algorithm for
finding exact minimal polynomial and reconstructing the exact root
by approximate value. Our algorithm is based on the improved
parameterized integer relation construction algorithm, whose
stability admits an efficient implementation with lower run times on
average than the former algorithm, and can be used to prove that
relation bounds obtained from computer runs using it are numerically
accurate. The other function \emph{identify} in \emph{maple} , which
finds a closed form for a decimal approximation of a number, was
implemented using the integer relation construction algorithm.
However, the choice of \emph{Digits} of approximate value is fairly
arbitrary \cite{BHM2003}. In contrast, we fully analyze numerical
behavior of an approximate to exact value and give how many
\emph{Digits} of approximate value, which can be obtained exact
results. The work is regard as a further research in \cite{ZF2007}.
We solve the problem, which can be described as follows:
\par Given approximate value $\tilde{\alpha}$ at arbitrary accuracy of
an unknown algebraic number, and
 we also know the degree of the algebraic number $n$ and an upper bound $N$ of
 its height on minimal polynomial in advance.
 The problem will be solved in two steps: First, we discuss how much
control error $\varepsilon$ is, so that we can reconstruct the
 algebraic  number $\alpha$ from its approximation
$\tilde{\alpha}$  when it holds that
$|\alpha-\tilde{\alpha}|<\varepsilon$. Of course, $\varepsilon$ is a
function in $n$ and $N$. Second, we give an algorithm to compute the
minimal polynomial of the algebraic number.
\par We are able to extend our results with the same methods to devise
a simple polynomial-time algorithm to factor multivariate
polynomials with rational coefficients, and provide a natural,
efficient technique to the minimal polynomial representation.
\par The rest of this paper is organized as follows. Section 2
illustrates the improved parameterized integer relation construction
algorithm. In Section 3, we discuss how to recover a quadratic
algebraic number and reconstruct minimal polynomial by
approximation. Section 4 gives some experimental results. The final
section concludes this paper.
\section{Preliminaries}
\label{sec:2}
\par In this section, we first give some notations,
and a brief introduction on integer relation problems. Then an
improved parameterized integer relation construction algorithm is
also reviewed.
\subsection{Notations}
 Throughout this paper, $\mathbf{Z}$ denotes the set of the integers,
$\mathbf{Q}$ the set of the rationals, $\mathbf{R}$ the set of the
reals, $\mathbb{O}(\mathbb{R}^{n})$ the corresponding system of
ordinary integers, $U(n-1,R)$ the group of unitary matrices over
$\mathbf{R}$, $GL(n,\mathbb{O}(\mathbb{R}))$ the group of unimodular
matrices with entries in the integers, $col_{i}$B the i-th column of
the matrix B. The ring of polynomials with integral coefficients
will be denoted $\mathbf{Z}[X]$. The $content$ of a polynomial
$p(X)$ in $\mathbf{Z}[X]$ is the greatest common divisor of its
coefficients. A polynomial in $\mathbf{Z}[X]$ is $primitive$ if its
content is 1. A polynomial $p(X)$ has degree $d$ if
$p(X)=\sum_{i=0}^{d}p_{i}X^{i}$ with $p_{d}\neq 0$. We write
$deg(p)=d$. The $length$ $|p|$ of $p(X)=\sum_{i=0}^dp_{i}X^{i}$ is
the Euclidean length of the vector $(p_{0},p_{1},\cdots,p_{d})$; the
$height$ $|p|_{\infty}$ of $p(X)$ is the $L_{\infty}$-norm of the
vector$(p_{0},p_{1},\cdots,p_{d})$, so $|p|_{\infty}=\max_{0\leq i
\leq d}|p_{i}|$. An $algebraic$ $number$ is a root of a polynomial
with integral coefficients. The $minimal$ $polynomial$ of an
algebraic number $\alpha$ is the irreducible polynomial in
$\mathbf{Z}[X]$ satisfied by $\alpha$. The minimal polynomial is
unique up to units in $\mathbf{Z}$. The $degree$ and $height$ of an
algebraic number are the degree and height, respectively, of its
minimal polynomial.
\subsection{Integer relation algorithm}
  There exists an integer relation amongst the numbers $x_{1}, x_{2},
\cdots, x_{n}$ if there are integers $a_{1}, a_{2}, \cdots, a_{n}$,
not all zero, such that $\sum_{i=1}^n a_{i}x_{i}=0$. For the vector
$\textbf{x}=[x_{1}, x_{2},\cdots, x_{n}]^{T}$, the nonzero vector
$\mathrm{a}\in\mathbb{Z}^{n}$ is an integer relation for
$\mathrm{x}$ if $\textbf{a}\cdot\textbf{x}=0$. Here are some useful
definitions and theorems\cite{FBA1999,BL2000}:
\begin{definition}($M_{x}$)
Assume ${x}=[x_{1}, x_{2},\cdots, x_{n}]^{T}\in \mathbb{R}^{n}$ has
norm $|x|$=1. Define $x^{\bot}$ to be the set of all vectors in
$\mathbb{R}^{n}$ orthogonal to $x$. Let
$\mathbb{O}(\mathbb{R}^{n})\cap x^{\bot}$be the discrete lattice of
integral relations for $x$. Define $M_{x}>0$ to be the smallest norm
of any relation for x in this lattice.
\end{definition}
\begin{definition}\label{def:Hx}
($H_{x}$) Assume ${x}=[x_{1}, x_{2},\cdots, x_{n}]^{T}\in
\mathbb{R}^{n}$ has norm $|x|$=1. Furthermore, suppose that no
coordinate entry of $x$ is zero, i.e., $x_{j}\neq 0$ for $1\leq
j\leq n$(otherwise $x$ has an immediate and obvious integral
relation). For $1\leq j\leq n$ define the partial sums
\begin{eqnarray*}
  s^{2}_{j}=\sum_{j\leq k \leq n}x_{k}^{2}.
\end{eqnarray*}
Given such a unit vector $x$, define the $n\times (n-1)$ lower
trapezoidal matrix $H_{x}=(h_{i,j})$ by
\par $$ h_{i,j}=
\begin{cases}
         0             \ \ \ \  \ \ \ \ \ \ \ \ \  \ \ \ \ \ \ \ \ \mbox{if $1\leq i<j\leq n-1,$}\\
         s_{i+1}/s_{i}   \ \ \ \ \ \ \ \ \ \ \ \ \ \  \mbox{if $1\leq i=j \leq n-1,$}\\
        -x_{i}x_{j}/(s_{j}s_{j+1})   \ \ \ \ \mbox{if $1 \leq j<i\leq
        n.$}\\
        \end{cases}
$$
Note that $h_{i,j}$ is scale invariant.
\end{definition}
\begin{definition}\label{def:ModifHR}
(Modified Hermite reduction) Let H be a lower trapezoidal matrix,
with $h_{i,j}=0$ if $j>i$ and $h_{j,j}\neq 0$. Set $D=I_{n}$, define
the matrix $D=(d_{i,j})\in GL(n,\mathbb{O}(R))$ recursively as
follows: For i from 2 to n, and for j from i-1 to 1(step-1), set
$q=nint(h_{i,j}/h_{j,j})$; then for k from 1 to j replace $h_{i,k}$
by $h_{i,k}-qh_{j,k}$, and for k from 1 to n replace
$d_{i,k}-qd_{j,k}$, where the function nint denotes a nearest
integer function, e.g., nint(t)=$\lfloor t+1/2 \rfloor$.
\end{definition}
\begin{theorem}\label{theo:the lower bound}
Let $x\neq 0\in \mathbb{R}^n$. Suppose that for any relation m of x
and for any matrix A $\in GL(n,\mathbb{O}(\mathbb{R}))$ there exists
a unitary matrix Q$\in$ U(n-1) such that $H = AH_{x}Q$ is lower
trapezoidal and all of the diagonal elements of H satisfy
$h_{j,j}\neq 0$. Then
 \begin{eqnarray*}
  \frac{1}{\max_{1\leq j \leq n-1
|h_{j,j}|}}=\min_{1\leq j \leq n-1}\frac{1}{|h_{j,j}|}\leq {|m|}.
\end{eqnarray*}
\end{theorem}
\begin{remark}
The inequality of Theorem \ref{theo:the lower bound} offers an
increasing lower bound on the size of any possible relation. Theorem
\ref{theo:the lower bound} can be used with any algorithm that
produces $GL(n,\mathbb{O}(\mathbb{R}))$ matrices. Any
$GL(n,\mathbb{O}(\mathbb{R}))$ matrix $A$ whatsoever can be put into
Theorem \ref{theo:the lower bound}.
\end{remark}
\begin{theorem}\label{theo:iterations}
Assume real numbers, $n\geq 2$, $\tau >1$, $\gamma>\sqrt{4/3}$, and
that $0\neq x\in \mathbb{R}^{n}$ has $\mathbb{O}(\mathbb{R})$
integer relations. Let $M_{x}$ be the least norm of relations for
$x$. Then $PSLQ(\tau)$ will find some integer relation for $x$ in no
more than
 \begin{eqnarray*}
 {n\choose 2 }\frac{log(\gamma^{n-1}M_{x})}{log\tau}
 \end{eqnarray*}
 iterations.
\end{theorem}
\begin{theorem}\label{theo:the upper bound}
Let $M_{x}$ be the smallest possible norm of any relation for x. Let
m be any relation found by PSLQ($\tau$). For all $\gamma >
\sqrt{4/3}$ for real vectors
 \begin{eqnarray*}
  |m| \leq \gamma^{n-2}M_{x}.
 \end{eqnarray*}
\end{theorem}
\begin{remark}
For n=2, Theorem \ref{theo:the upper bound} proves that any relation
$0\neq m \in \mathbb{O}(\mathbb{R}^{2})$ found has norm $|m|=M_{x}$.
In other words, $PSLQ(\tau)$ finds a shortest relation. For real
numbers this corresponds to the case of the Euclidean algorithm.
\end{remark}
\par Based on the theorems as above, and if there exists a known error controlling $\varepsilon$,
then an algorithm for obtaining the
integer relation was designed as follows:
\newcounter{num}
\begin{algorithm}\label{alg:integer relation}Parameterized
integer relation construction algorithm\\
Input: a vector $x$, and an error control $\varepsilon>0$;\\
Output: an integer relation $m$.
\begin{list}{Step \arabic{num}:}{\usecounter{num}\setlength{\rightmargin}{\leftmargin}}
\item Set $i:=1$,  $\tau:=2/\sqrt3$, and unitize the vector $x$ to $\bar{x}$;
\item Set $H_{\bar{x}}$ by definition \ref{def:Hx};
\item Produce matrix $D \in GL(n, \mathbb{O}(\mathbb{R}))$ using modified Hermite
Reduction;
\item Set $\bar{x}:=\bar{x}\cdot D^{-1}, H:=D\cdot H, A:=D\cdot A, B:=B\cdot
D^{-1}$, \\
case 1: if $\bar{x}_{j}=0$, then $m:=col_{j}B$, goto Step 11; \\
case 2: if $h_{i,i}<\varepsilon$, then $m:=col_{n-1}B$, goto Step
11;
\item $i:=i+1$;
\item Choose an integer r, such that $\tau^{r}|h_{r,r}|\geq
\tau^{j}|h_{i,i}|$, for all $1\leq j \leq n-1$;
\item Define $\alpha:=h_{r,r}$, $\beta:= h_{r+1,r}$,
$\lambda:=h_{r+1,r+1}$, $\sigma:=\sqrt{\beta^2+\lambda^2}$;
\item Change $h_{r}$ to $h_{r+1}$, and define the permutation matrix R;
\item Set $\bar{x}:=\bar{x}\cdot R$, $H:= R\cdot H$, $A:=R\cdot A$,
$B:= B\cdot R$, if i=n-1, then goto Step 4;
\item Define $ Q:=(q_{i,j})\in U(n-1,R)$, $H:=H \cdot Q$, goto Step 4;
\item return $m$.
\end{list}
\end{algorithm}
\par By algorithm \ref{alg:integer relation}, we can find the integer
relation $m$ of the vector
$x=(1,\tilde{\alpha},\tilde{\alpha}_{2},\cdots,\tilde{\alpha}_{n})$.
So, we get a nonzero polynomial of degree $n$, i.e.,
\begin{equation}\label{equ:polynomial}
G(x)=m\cdot(1,x,x^{2},\cdots,x^{n})^{T}.
\end{equation}
\par Our main task is to show that polynomial (\ref{equ:polynomial})
is uniquely determined under assumptions, and discuss the
controlling error $\varepsilon$ in algorithm \ref{alg:integer
relation} in the next section.

\section{Reconstructing minimal polynomial from its approximation}
\par In this section, we will solve such a problem: For a given
floating number $\tilde{\alpha}$, which is an approximation of
unknown algebraic number, how do we obtain the exact value? Without
loss of generality, we first consider the recovering quadratic
algebraic number from its approximate value, and then generalize the
results to the case of algebraic number of high degree. At first, we
have some lemmas as follows:
\begin{lemma}\label{lem:errorctrol}
Let $f=\sum_{i=0}^{n}a_{i}x^{i}\in \mathbb{Z}[x]$ be a polynomial of
degree $n>0$, and let $\varepsilon=\max_{1\leq i\leq
n}|\alpha^{i}-\tilde{\alpha_{i}}|$ for the rest of this paper, where
$\tilde{\alpha}_{i}$ for $1\leq i \leq n$ are the rational
approximations to the powers $\alpha^{i}$ of algebraic number
$\alpha$, and $\tilde{\alpha}_{0}=1$. Then
\begin{equation}\label{equ:errorctrol}
     |f(\alpha)-f(\tilde{\alpha})|\leq \varepsilon \cdot n \cdot
     |f|_{\infty}.
\end{equation}
\end{lemma}
{\bf Proof:} Since
$f(\alpha)-f(\tilde{\alpha})=\sum_{i=0}^{n}a_{i}(\alpha^{i}-\tilde{\alpha}_{i})$,
we get
$|f(\alpha)-f(\tilde{\alpha})|=|\sum_{i=1}^{n}a_{i}(\alpha^{i}-\tilde{\alpha}_{i})|$,
and then
\begin{eqnarray*}
|\sum_{i=1}^{n}a_{i}(\alpha^{i}-\tilde{\alpha}_{i})|\leq
\sum_{i=1}^{n}|a_{i}|\cdot|(\alpha^{i}-\tilde{\alpha}_{i})|\leq
\sum_{i=1}^{n}|a_{i}|\cdot \varepsilon \leq n\cdot |f|_{\infty}\cdot
\varepsilon.
\end{eqnarray*}
So, the lemma is finished.
\begin{lemma}\label{lem:belowbound}
Let $h$ and $g$ be nonzero polynomials in $\mathbb{Z}[x]$ of degree
$n$ and $m$, respectively, and let $\alpha\in \mathbb{R}$ be a zero
of $h$ with $|\alpha|\leq 1$. If $h$ is irreducible and
$g(\alpha)\neq 0$, then
\begin{equation}\label{equ:belowbound}
     |g(\alpha)|\geq n^{-1}\cdot |h|^{-m}\cdot |g|^{1-n}.
\end{equation}
\end{lemma}
{\bf Proof:} See Proposition(1.6) of\cite{KLL1988}. If $|\alpha|>
1$, a simple transform of it does.
\begin{corollary}\label{cor:extendbelowbound}
Let $h$ and $g$ be nonzero polynomials in $\mathbb{Z}[x]$ of degrees
$n$ and $m$, respectively, and let $\alpha\in \mathbb{R}$ be a zero
of $h$ with $|\alpha|\leq 1$. If $h$ is irreducible and
$g(\alpha)\neq 0$, then
\begin{equation}\label{equ:belowbound}
     |g(\alpha)|\geq n^{-1}\cdot (n+1)^{-\frac{m}{2}}\cdot (m+1)^{\frac{1-n}{2}}\cdot |h|^{-m}_{\infty}\cdot |g|^{1-n}_{\infty}.
\end{equation}
\end{corollary}
{\bf Proof:} First notice that $|f|^{2}\leq (n+1)\cdot
|f|^{2}_{\infty}$ holds for any polynomial $f$ of degree at most
$n>0$, so $|f|\leq \sqrt{n+1}\cdot |f|_{\infty}$. From Lemma
\ref{lem:belowbound} we get
\begin{eqnarray*}
     |g(\alpha)|\geq n^{-1}\cdot (n+1)^{-\frac{m}{2}}\cdot (m+1)^{\frac{1-n}{2}}\cdot |h|^{-m}_{\infty}\cdot |g|^{1-n}_{\infty}.
\end{eqnarray*}
\par So, the corollary is finished.
\begin{theorem}\label{theo:the minimal polynomial}
Let an approximate value $\tilde{\alpha}$ belong to an unknown
algebraic number $\alpha$ of degree $n>0$. Assume that the existence
of the polynomial $G(x)=\sum_{i=0}^{n}{a_{i}}x^{i}$, where
$a_{n}\neq 0$. Suppose $n$ and upper bound $N$ on the degree and
height of minimal polynomial $g(x)$ on the algebraic number $\alpha$
are known, respectively. If
\begin{eqnarray*}
|G(\tilde{\alpha})|<n^{-1}\cdot (n+1)^{-n+\frac{1}{2}}\cdot
|G|^{-n}_{\infty}\cdot N^{1-n}-n\cdot\varepsilon \cdot |G|_{\infty},
\end{eqnarray*}
then
\begin{eqnarray*}
|G(\alpha)|<n^{-1}\cdot (n+1)^{-n+\frac{1}{2}}\cdot
|G|^{-n}_{\infty}\cdot N^{1-n}.
\end{eqnarray*}
\end{theorem}
{\bf Proof:} From lemma \ref{lem:errorctrol}, we notice that
$|G(\alpha)-G(\tilde{\alpha})|\leq \varepsilon \cdot n \cdot
|G|_{\infty}$, and
\begin{eqnarray*}
|G(\alpha)-G(\tilde{\alpha})|\geq |G(\alpha)|-|G(\tilde{\alpha})|,\
so \ |G(\alpha)|\leq|G(\tilde{\alpha})|+n\cdot \varepsilon \cdot
|G|_{\infty}.
\end{eqnarray*}
From the assumption of the theorem, since
\begin{equation}\label{equ:max value}
|G(\tilde{\alpha})|<n^{-1}\cdot (n+1)^{-n+\frac{1}{2}}\cdot
|G|^{-n}_{\infty}\cdot N^{1-n}-n\cdot\varepsilon \cdot |G|_{\infty}.
\end{equation}
So, the theory is finished.
\begin{corollary}\label{cor:minimalpoly}
If $|G(\alpha)|<n^{-1}\cdot (n+1)^{-n+\frac{1}{2}}\cdot
|G|^{-n}_{\infty}\cdot N^{1-n}$, where $G(x)$ is constructed by the
parameterized integer relation construction algorithm as above, the
upper bound $N$ on the height of its minimal polynomial $g(x)$ on an
algebraic number $\alpha$ are known. Then
\begin{equation}\label{equ:belowbound}
    G(\alpha)=0,
\end{equation}
and the primitive part of  polynomial $G(x)$ is the minimal
polynomial of algebraic number $\alpha$.
\end{corollary}
{\bf Proof:} Proof is given by contradiction. According to Lemma
\ref{lem:belowbound}, suppose on the contrary that $G(\alpha)\neq
0$, then
$$|G(\alpha)|\geq n^{-1}\cdot (n+1)^{-n+\frac{1}{2}}\cdot
|G|^{-n}_{\infty}\cdot N^{1-n}.$$
 From theory \ref{theo:the minimal
polynomial}, we get
$$|G(\alpha)|<n^{-1}\cdot
(n+1)^{-n+\frac{1}{2}}\cdot |G|^{-n}_{\infty}\cdot N^{1-n}.$$ So,
$G(\alpha)=0$. Since algebraic number $\alpha$ is degree $n>0$, then
the primitive polynomial of $G(x)$ denotes $pp(G(x))$, hence
$pp(G(x))$ is just irreducible and equal to $g(x)$. Of course, it is
unique.
\par So, the corollary is finished.
\subsection{Recovering  quadratic algebraic number from approximate value}
\par For simplicity, we discuss how to obtain quadratic algebraic
number from its approximation using integer relation algorithm. Let
$\tilde{\alpha}$ be the approximate value, considering the vector
$v=(1, \tilde{\alpha}, \tilde{\alpha}^2)$. Our goal is to find a
vector $w$ which has all integer entries such that the dot product
of $v$ and $w$ is less than a lower bound, which is obtained and we
are able to get  the size of the neighborhood is
$1/(12\sqrt{3}N^{4})$ from theorem \ref{theo:the minimal
polynomial}. The following theorem answers the basic questions of
this approach.
\begin{theorem}\label{theo:the quadratic algebraic number}
Let $\tilde{\alpha}$ be an approximate value belonging to an unknown
quadratic algebraic number $\alpha$, if
\begin{equation}\label{equ:maxmin}
\varepsilon=|\alpha-\tilde{\alpha}|<1/(12\sqrt{3}N^{4}),
\end{equation}
where $N$ is the upper bound on the height of its minimal
polynomial. Then $G(\alpha)=0$, and the primitive part of $G(x)$ is
its minimal polynomial, where $G(x)=\sum_{i=0}^{2}a_{i}x^{i}$ is
constructed using integer relation algorithm as above.
\end{theorem}
{\bf Proof:} From theorem \ref{theo:the minimal polynomial} and
corollary \ref{cor:minimalpoly}, if and only if
$$|G(\alpha)|<n^{-1}\cdot (n+1)^{-n+\frac{1}{2}}\cdot
|G|^{-n}_{\infty}\cdot N^{1-n},$$ Therefore, $G(\alpha)=0$. Under
the assumption of the theorem we get $n=2$ and
$|G(\tilde{\alpha})|>0$, hence it is obvious that inequality
(\ref{equ:max value})holds, i.e.,
\begin{equation}\label{equ:inequalitysolve}
0<n^{-1}\cdot(n+1)^{-n+\frac{1}{2}}\cdot |G|^{-n}_{\infty}\cdot
N^{1-n}-n\cdot\varepsilon \cdot |G|_{\infty}.
\end{equation}
So, Solving inequality (\ref{equ:inequalitysolve}) yields
\begin{eqnarray*}
\varepsilon<1/(12\sqrt{3}|G|_{\infty}^{3}N).
\end{eqnarray*}
From theorem \ref{theo:the upper bound}, and $\alpha$ is a quadratic
algebraic number, so $|G|_{\infty}$ is just equal to $N$.
\par So, the theory is finished.
\par This theorem leads to the following algorithm for recovering
the quadratic algebraic number of $\tilde{\alpha}$:
\newcounter{num1}
\begin{algorithm}\label{alg:quadratic algebraic}Recovering quadratic algebraic number algorithm\\
Input: an floating number ($\tilde{\alpha}$, $N$) belonging to an unknown quadratic algebraic number $\alpha$,
i.e., satisfying (\ref{equ:maxmin}).\\
Output: an quadratic algebraic number $\alpha$.
\begin{list}{Step \arabic{num}:}{\usecounter{num}\setlength{\rightmargin}{\leftmargin}}
\item Construct the vector $v$;
\item Compute $\varepsilon$ satisfying (\ref{equ:maxmin});
\item Call algorithm \ref{alg:integer relation} to find an integer
relation $w$ for $v$;
\item Obtain $w(x)$ the corresponding polynomial;
\item Let $g(x)$ be the primitive part of $w(x)$;
\item Solve the equation $g(x)=0$ and choose the corresponding algebraic number
to $\alpha$;
\item return $\alpha$.
\end{list}
\end{algorithm}
\begin{theorem}
\par Algorithm \ref{alg:quadratic algebraic} works correctly as
specified and uses O($logN$) binary bit operations, where $N$ is the
upper bound of height on its minimal polynomial.
\end{theorem}
{\bf Proof:} Correctness follows from theorem \ref{theo:the
quadratic algebraic number}. The cost of the algorithm is O($logN$)
binary bit operations obviously.
\subsection{Obtaining minimal polynomial of high degree}
If $\alpha$ is a real number, then by definitioin $\alpha$ is
algebraic exactly if, for some $n$, the vector
\begin{equation}\label{equ:the vector}
(1,\alpha,\alpha^{2},\cdots,\alpha^{n})
\end{equation}
has an integer relation. The integer coefficient polynomial of
lowest degree, whose root $\alpha$ is, is determined uniquely up to
a constant multiple; it is called the $minimal polynomial$ for
$\alpha$. Integer relation algorithm can be employed to search for
minimal polynomial in a straightforward way by simply feeding them
the vector (\ref{equ:the vector}) as their input. Let
$\tilde{\alpha}$ be an approximate value belonging to an unknown
algebraic number $\alpha$, considering the vector
$v=(1,\tilde{\alpha},\tilde{\alpha}^{2},\cdots,\tilde{\alpha}^{n})$,
how to obtain the exact minimal polynomial from its approximate
value? We have the same technique answer to the question from the
following theorem.
\begin{theorem}\label{theo:the n-th algebraic number }
Let $\tilde{\alpha}$ be an approximate value belonging to an unknown
algebraic number $\alpha$ of degree $n>0$, if
\begin{equation}\label{equ:n-th maxmin}
\varepsilon=|\alpha-\tilde{\alpha}|<1/(n^{2}(n+1)^{n-\frac{1}{2}}N^{2n}),
\end{equation}
where $N$ is the upper bound on the height of its minimal
polynomial. Then $G(\alpha)=0$, and the primitive part of $G(x)$ is
its minimal polynomial, where $G(x)=\sum_{i=0}^{n}a_{i}x^{i}$ is
constructed using the parameterized integer relation construction
algorithm as above.
\end{theorem}
{\bf Proof:} The proof can be given similarly to that in theorem
\ref{theo:the quadratic algebraic number}.
\par It is easiest to appreciate the theorem by seeing how it
justifies the following algorithm for obtaining minimal polynomials
from its approximation:
\begin{algorithm}\label{alg:n-th algebraic}Obtaining minimal polynomial algorithm\\
Input: an floating number ($\tilde{\alpha}$, $n$, $N$) belong to an unknown  algebraic number $\alpha$,
i.e., satisfying (\ref{equ:n-th maxmin}).\\
Output: $g(x)$, the minimal polynomial of $\alpha$.
\begin{list}{Step \arabic{num}:}{\usecounter{num}\setlength{\rightmargin}{\leftmargin}}
\item Construct the vector $v$;
\item Compute $\varepsilon$ satisfying (\ref{equ:n-th maxmin});
\item Call algorithm \ref{alg:integer relation} to find an integer
relation $w$ for $v$;
\item Obtain $w(x)$ the corresponding polynomial;
\item Let $g(x)$ be the primitive part of $w(x)$;
\item return $g(x)$.
\end{list}
\end{algorithm}
\begin{theorem}
\par Algorithm \ref{alg:n-th algebraic} works correctly as specified
and uses O($n(logn+logN)$) binary bit operations, where  $n$ and $N$
are the degree and height of its minimal polynomial, respectively.
\end{theorem}
{\bf Proof:} Correctness follows from theorem \ref{theo:the n-th
algebraic number }. The cost of the algorithm is O($n(logn+logN)$)
binary bit operations obviously.
\par The method of obtaining minimal polynomials from an
approximate value can be extended to the set of complex numbers and
many applications in computer algebra and science.
\par This yields a simple factorization algorithm for multivariate polynomials with
rational coefficients: We can reduce a multivariate polynomial to a
bivariate polynomial using the Hilbert irreducibility theorem, the
basic idea was described in \cite{CAW2002}, and then convert a
bivariate polynomial to a univariate polynomial by substituting a
transcendental number in \cite{MA1985} or an algebraic number of
high degree for a variate in \cite{CFQZ2008}. It can find the
bivariate polynomial's factors, from which the factors of the
original multivariate polynomial can be recovered using Hensel
lifting. After this substitution we can get an approximate root of
the univariate polynomial and use our algorithm to find the
irreducible polynomial satisfied by the exact root, which must then
be a factor of the given polynomial. This is repeated until the
factors are found.
\par The other yields an efficient method of converting the rational
approximation representation to the minimal polynomial
representation. The traditional representation of algebraic numbers
is by their minimal polynomials
\cite{BC1990,BCRD1986,EP1997,RL1982}. We now propose an efficient
method to the minimal polynomial representation, which only needs an
approximate value, degree and height of its minimal polynomial,
i.e., an ordered triple $<\tilde{\alpha},n,N>$ instead of an
algebraic number $\alpha$, where $\tilde{\alpha}$ is its approximate
value, and $n$ and $N$ are the degree and height of its minimal
polynomial, respectively, denotes $<\alpha>=<\tilde{\alpha},n,N>$.
It is not hard to see the computations in the representation can be
changed to computations in the other without loss of efficiency, the
rational approximation method is closer to the intuitive notion of
computation.
\section{Experimental Results}
Our algorithms are implemented in \emph{Maple}. The following
examples run in the platform of Maple 11 and PIV3.0G,512M RAW. The
first three examples illuminate how to obtain exact quadratic
algebraic number and minimal polynomials. Example 4 tests our
algorithm for factoring primitive polynomials with integral
coefficients.
\par \textbf{Example 1}  Let $\alpha$ be an unknown
quadratic algebraic number. We only know an upper bound of height on
its minimal polynomial $N=47$. According to theorem \ref{theo:the
quadratic algebraic number}, compute quadratic algebraic number
$\alpha$ as follows: First obtain control error
$\varepsilon=1/(12*\sqrt{3}*N^{4})=1/(1807729447692*\sqrt{3})\approx
1.0\times10^{-8}$. And then assume that we use some numerical method
to get an approximation $\tilde{\alpha}=11.937253933$, such that
$|\alpha-\tilde{\alpha}|<\varepsilon$. Calling algorithm
\ref{alg:quadratic algebraic} yields as follows: \\
Its minimal polynomial is $g(x)=x^{2}-8*x-47$. So, we can obtain the
corresponding quadratic algebraic number $\alpha=4+3\sqrt{7}$.
\begin{remark}
  The function \emph{identify} in maple 11 needs \emph{Digits}=13,
  whereas our algorithm only needs 9 digits.
\end{remark}
\par  \textbf{Example 2}  For obtaining exact minimal polynomials
from approximate root $\tilde{\alpha}$, we only know degree $n=3$
and height $N=17$ of its minimal polynomial. According to theorem
\ref{theo:the n-th algebraic number }, just as do in Example 1:
First get the error
$\varepsilon=1/(n^{2}(n+1)^{n-\frac{1}{2}}N^{2n})=1/6951619872\approx1.4\times10^{-10}$.
Assume that we use some numerical method to get an approximation
$\tilde{\alpha}=16.808034642702$, such that
$|\alpha-\tilde{\alpha}|<\varepsilon$. Calling algorithm
\ref{alg:n-th algebraic} yields as follows: Its minimal polynomial
is $g(x)=x^{3}-17*x^{2}+4*x-13$.
\par \textbf{Example 3} Let a known floating number $\tilde{\alpha}$ belonging
to some algebraic number $\alpha$ of degree $n=4$, where
$\tilde{\alpha}=3.14626436994198$, we also know an upper bound of
height $N=10$ on its minimal polynomial. According to theorem
\ref{theo:the n-th algebraic number }, we can get the error
$\varepsilon=1/(n^{2}(n+1)^{n-\frac{1}{2}}N^{2n})=1/(4^{2}*5^{\frac{7}{2}}*10^{8})\approx2.2\times10^{-12}$.
Calling algorithm \ref{alg:n-th algebraic}, if only the floating
number $\tilde{\alpha}$, such that
$|\alpha-\tilde{\alpha}|<\varepsilon$, then we can get its minimal
polynomial $g(x)=x^{4}-10*x^{2}+1$. So, the exact algebraic number
$\alpha$ is able to denote $<\alpha>=<3.14626436994198,4,10>$, i.e.,
$<\sqrt{2}+\sqrt{3}>=<3.14626436994198,4,10>$.
\par \textbf{Example 4} This example is an application in factoring primitive polynomials over integral coefficients.
For the conveniency of display in the paper, we choose a very simple
polynomial as follows:
$$p=3x^{9}-9x^{8}+3x^{7}+6x^{5}-27x^{4}+21x^{3}+30x^{2}-21x+3$$
We want to factor the polynomial $p$ via reconstruction of minimal
polynomials over the integers. First, we transform $p$ to a
primitive polynomial as follows:
$$p=x^{9}-3x^{8}+x^{7}+2x^{5}-9x^{4}+7x^{3}+10x^{2}-7x+1,$$
We see the upper bound of coefficients on polynomial $p$ is $10$,
which has relation with an upper bound of coefficients of the
factors on the primitive polynomial $p$ by Landau-Mignotte bound
\cite{M1974}. Taking $N=5$, $n=2$ yields
$\varepsilon=1/(2^{2}*(2+1)^{2-\frac{1}{2}}*5^{4})=1/(7500*\sqrt{3})\approx8.0\times10^{-5}$.
Then we compute the approximate root on $x$. With Maple we get via
[fsolve($p=0,x$)]:
  $$S=[2.618033989, 1.250523220, -.9223475138, .3819660113, .2192284350]$$
According to theorem \ref{theo:the n-th algebraic number }, let
$\tilde{\alpha}=2.618033989$ be an approximate value belonging to
some quadratic algebraic number $\alpha$, calling algorithm
\ref{alg:n-th algebraic} yields as follows: $$p_{1}=x^{2}-3*x+1.$$
And then we use the polynomial division to get
$$p_{2}=x^{7}+2*x^{3}-3*x^{2}-4*x+1.$$
Based on the Eisenstein's Criterion \cite{SL2002}, the $p_{2}$ is
irreducible in $\mathbf{Z}[X]$. So, the $p_{1}$ and $p_{2}$ are the
factors of primitive polynomial $p$.

\section{Conclusion}
\par In this paper, we have presented a new method for obtaining exact
results by numerical approximate computations. The key technique of
our method is based on an improved parameterized integer relation
construction algorithm, which is able to find an exact relation by
the accuracy control $\varepsilon$ in formula (\ref{equ:n-th
maxmin}) is an exponential function in degree and height of its
minimal polynomial. The result is almost consistent with the
existence of error controlling on obtaining an exact rational number
from its approximation in \cite{ZF2007}. Using our algorithm, we
have succeed in factoring multivariate polynomials with rational
coefficients and providing an efficient method of converting the
rational approximation representation to the minimal polynomial
representation. Our method can be applied in many aspects, such as
proving inequality statements and equality statements, and computing
resultants, etc.. Thus we can take fully advantage of approximate
methods to solve larger scale symbolic computation problems.

\end{document}